\documentclass[reprint,amsmath,amssymb,aps]{revtex4-2}

\usepackage[pdftex]{graphicx}
\usepackage{dcolumn}
\usepackage{bm}
\usepackage{amsmath}
\usepackage[usenames,dvipsnames]{xcolor}
\usepackage{mathtools}
\usepackage[colorlinks=true,citecolor=blue,linkcolor=blue]{hyperref}
\usepackage{bbm}
\usepackage{comment}
\usepackage{upgreek}
\usepackage{esint}
\usepackage{soul}
\usepackage[normalem]{ ulem } 
\usepackage{floatrow}
\usepackage{verbatim}
\usepackage{braket}
\usepackage{amsmath, amssymb}
\usepackage{array}
\usepackage{algorithm}
\usepackage{algpseudocode}
\usepackage{graphicx}
\usepackage{xcolor} % put this in your preamble
\usepackage{enumitem}
\usepackage{subcaption}
\usepackage{multirow}
\usepackage{booktabs}
\usepackage{array} % preamble

\usepackage[colorlinks=true,citecolor=blue,linkcolor=blue]{hyperref}
\setlist[itemize]{parsep=1em, itemsep=1em} % Adjust spacing for itemize
\setlist[enumerate]{parsep=0em, itemsep=1em} % Adjust spacing for enumerate

\newcommand{\PreserveBackslash}[1]{\let\temp=\\#1\let\\=\temp}
\newcolumntype{C}[1]{>{\PreserveBackslash\centering}p{#1}}

\newcommand{\rom}[1]{\uppercase\expandafter{\romannumeral #1\relax}}
\newcommand{\umqz}{U_\text{MQ}^\text{(Z)} }
\newcommand{\umqx}{U_\text{MQ}^\text{(X)} }
\newcommand{\umqp}{U_\text{MQ}^\text{(P)} }

\setcitestyle{numbers,sort&compress}

\begin{document}

\title{Optimal constant-cost implementations of Clifford operations using global interactions}

\author{Jonathan Nemirovsky}
\author{Lee Peleg}
\author{Amit Ben Kish}
\author{Yotam Shapira}\email{yotam.shapira@quantum-art.tech}

\affiliation{Quantum Art, Ness Ziona 7403682, Israel}

\begin{abstract}
	We investigate quantum circuits built from arbitrary single-qubit operations combined with programmable all-to-all multiqubit entangling gates that are native to, among other systems, trapped-ion quantum computing platforms. We report a constant-cost of no more than four applications of such Clifford entangling multiqubit gates to realize any sequence of Clifford operations of any length, without ancillae, which is the theoretically optimal gate count cost. We do this by implementing any sequence of CNOT gates of any length with four applications of such gates, without ancillae, and show that the extension to general Clifford operations incurs no additional cost. We investigate the required qubit drive power that is associated with our implementation and show that it is lower than that of a standard approach. Our work introduces a practical and computationally efficient algorithm to realize these compilations.	
\end{abstract}

\maketitle

Clifford operations are central to many quantum information processing applications such as quantum error correction, simulation algorithms, generation of pseudo-random unitaries, qubit permutations, compilation of quantum circuits and benchmarking \cite{vandenberg2020circuit,haferkamp2023efficient,kueng2015qubit,zhu2017multiqubit,elben2023randomized,huang2022foundations,huang2020predicting,nemirovsky2025efficient,nemirovsky2025phasegadgetcompilationquantum}. Thus, it is of considerable interest to develop efficient decompositions of Clifford operations.

Here we provide an explicit algorithm that decomposes arbitrary Clifford operations into a quantum circuit that utilizes single-qubit rotations and, at most, four multiqubit programmable all-to-all entangling gates, which saturates the lower bound \cite{cleve2025improvedcliffordoperationsconstant}. Previous works have shown implementations of Clifford operations using such multiqubit gates with a gate count that is linear (in the number of qubit), logarithmic or constant \cite{maslov2018use,van2021constructing,maslov2022depth,bravyi2022constant,cleve2025improvedcliffordoperationsconstant}, with Ref.~\cite{bravyi2022constant} greatly inspiring this work.

Explicitly, the multiqubit Clifford gate over $n$ qubits we consider is,
\begin{equation}
	\umqp\left(\xi\right)=e^{-i\frac{\pi}{2}\sum_{k=1}^n \xi_{kk} P_k - i\frac{\pi}{4}\sum_{k>j}^n \xi_{kj} P_k P_j},
	\label{eqUmq}
\end{equation}
where $P_k$ is the $P\in\{X,Y,Z\}$ Pauli operator, acting on the $k$th qubit and $\xi\in\mathbb{F}_2^{n\times n}$ is an arbitrary symmetric matrix of zeros and ones, such that for $k \ne j$, $\frac{\pi}{4}\xi_{kj}$ is a correlated $P\otimes P$ rotation between the $k$th and $j$th qubits. We note that $\umqx$ and $\umqz$ are equivalent up to single-qubit Hadamard gates, yet in our derivations below it is simpler to use them both.

These gates can be economically implemented on trapped-ions based quantum computers \cite{feng2023continuous,pogorelov2021compact,yao2022experimental,guo2024site,schwerdt2024scalable,nemirovsky2025efficient}, where the long-range Coulomb coupling between the ions underpins the all-to-all programmability \cite{grzesiak2020efficient,shapira2020theory,shapira2023fast,lu2019global,bassler2023synthesis,lu2019global,lu2025implementing,shapira2025programmable}. Additional quantum computing modalities have the potential to generate such programmable multiqubit interactions \cite{evered2023high,jeremy2021asymmetric,cooper2024graph,yongxin2025constant}, and can as well benefit from our method.

Our decomposition relies on two steps. We first show that an arbitrary Clifford operation can be decomposed to the form (operating left to right),

\begin{equation}
	U_\text{C}=-\text{P}-\text{S}_Z-\text{H}_\text{all}-\text{R}_{XX}-\text{CX}-\text{R}_{ZZ}-\text{S}_Z-\text{H}-,\label{eqDecomp}
\end{equation} 
by a simple extension of the decompositions in Refs.~\cite{proctor2023simple,bravyi2022constant,bravyi2021hadamard,nemirovsky2025reducedconstantcostimplementationsclifford}, with $-\text{P}-$ a layer of Pauli gates, $-\text{S}_Z-$ a layer of single-qubit $Z^{1/2}$ phase gates, $-\text{H}_\text{all}-$ a layer of Hadamard gates over all qubits, $-\text{R}_{XX}-$ and $-\text{R}_{ZZ}-$ correlated rotation layers of $R_{XX}=\exp\left(i\frac{\pi}{4}X\otimes X\right)$ and $R_{ZZ}=\exp\left(i\frac{\pi}{4}Z\otimes Z\right)$ gates, respectively, $-\text{CX}-$ a classical linear reversible circuit made of CNOT gates and $-\text{H}-$ a layer of Hadamard gates. Clearly the $-\text{R}_{ZZ}-$ layer can be implemented with a single use of $\umqz$ and similarly the $-\text{R}_{XX}-$ with a single use of $\umqx$. 

Next, we show that the linear reversible circuit, $-\text{CX}-$, can be decomposed with at most four multiqubit gates. Moreover, this decomposition starts with a $\umqx$ gate and terminates with a $\umqz$ gate. These `edge' gates are then merged with the correlated rotation layers, such that sequence $-\text{R}_{ZZ}-\text{CX}-\text{R}_{XX}-$ from Eq. \eqref{eqDecomp} is implemented with, at most, four multiqubit gates. Since all other gates in the decomposition are single-qubit gates, then overall the Clifford operation itself is implemented with, at most, four multiqubit gates. This gate cost is the lower bound, as shown in Ref.~\cite{cleve2025improvedcliffordoperationsconstant}.

We work within the symplectic matrix formalism, in which Clifford operations over $n$ qubits are mapped in a homeomorphism to $2n\times2n$ matrices over the binaries $\mathbb{F}_2=\left\{0,1\right\}$~\cite{gottesman1997stabilizer,aaronson2004improved}. The symplectic matrix is equivalent, up to Pauli strings which are easy to compute, to the Clifford operation. 

Specifically, the symplectic forms of the $-\text{CX}-$ layer, $\umqz\left(\xi\right)$ and $\umqx\left(\xi\right)$ are given by,
\begin{align}
	S(\umqz(\xi)) =
	\begin{bmatrix}
		I_n & 0 \\
		\xi & I_n
	\end{bmatrix}, \label{eqSumqz}\\ 
	S(\umqx(\xi))=
	\begin{bmatrix}
		I_n & \xi \\
		0 & I_n
	\end{bmatrix},\label{eqSumqx}\\
		S\left(-\text{CX}-\right)=
	\begin{bmatrix}
		C^{-T} & 0 \\
		0 & C
	\end{bmatrix},\label{eqSCX}
\end{align}
with $\xi$ a symmetric matrix as in Eq.~\eqref{eqUmq}, $I_n$ the $n\times n$ identity operator and $C$ an invertible matrix such that $C^{-T}$ is its inverse transpose. These representation are explicitly derived in Ref.~\cite{nemirovsky2025reducedconstantcostimplementationsclifford}.

We first derive the general decomposition in Eq.~\eqref{eqDecomp}. We start from the known decomposition in Refs.~\cite{bravyi2021hadamard,proctor2023simple},
\begin{equation}
	U_C = -\text{P}-\text{CX}-\text{R}_\text{ZZ}-\text{S}_{Z}-\text{H}_\text{all}-\text{R}_\text{ZZ}  -\text{S}_\text{Z}-\text{H}-,\label{eqProctor}
\end{equation}
We proceed by commuting the $-\text{H}_\text{all}-$ layer towards the left-hand side. Since this layer operates with a Hadamard on all qubits it transforms $-\text{S}_{Z}-\mapsto-\text{S}_{X}-$, a set of $X^{1/2}$ single qubit rotations and similarly $-\text{R}_\text{ZZ}-\mapsto-\text{R}_\text{XX}-$. The $-\text{CX}-$ layer is modified such that the roles of target and control qubits are exchanged, retaining a $-\text{CX}-$ layer structure. We obtain,
\begin{equation}
	U_C = -\text{P}-\text{H}_\text{all}-\text{CX}-\text{R}_\text{XX}-\text{S}_\text{X}-\text{R}_\text{ZZ}  -\text{S}_\text{Z}-\text{H}-,\label{eqProctorCommute}
\end{equation}
In the symplectic representation, Eq.~\eqref{eqProctorCommute} becomes,
\begin{equation}
	\begin{split}
		S(U_C) =& S(-\text{H}-) S(-\text{R}_\text{ZZ}-\text{S}_\text{Z}-)\\ & S(-\text{R}_\text{XX}-\text{S}_\text{X}-)  S(-\text{CX}-) S(-\text{H}_\text{all}-) \label{eq:layer_rule}
	\end{split}
\end{equation}
Focusing on $\tilde{S}\equiv S(-\text{R}_\text{XX}-\text{S}_\text{X}-)  S(-\text{CX}-) $, we note that,
\begin{equation}
\begin{split}
	\tilde{S} & = 
	\begin{bmatrix}
		I & M \\
		0 & I
	\end{bmatrix}
	\begin{bmatrix}
		C^{-T} & 0 \\
		0 & C
	\end{bmatrix} 
	=
	\begin{bmatrix}
		C^{-T} & 0 \\
		0 & C
	\end{bmatrix}
	\begin{bmatrix}
		I & C^{T}M C \\
		0 & I
	\end{bmatrix} \\
	&= S(-\text{CX}-)  S(-\text{R}_\text{XX}-\text{S}_\text{X}-).
\end{split}\label{eqCommute}
\end{equation}
That is, the layers commute by modifying the $\text{R}_\text{XX}-$ and $-\text{S}_\text{X}-$ layers. We note that this fact is also intuitive by considering the operation of CNOT gates on phase-gadgets \cite{nemirovsky2025phasegadgetcompilationquantum}. Applying Eq.~\eqref{eqCommute} to Eq.~\eqref{eq:layer_rule}, and commuting the $-\text{S}_\text{X}-$ layer with $-\text{H}_\text{all}-$, we obtain the decomposition shown in Eq.~\eqref{eqDecomp} above, concluding the proof.

Next we show a decomposition of the $-\text{CX}-$ layer with at most four multiqubit gates that starts with $\umqx$ and terminates with $\umqz$. We recall that the symplectic matrix of the $-\text{CX}-$ layer, $S\left(-\text{CX}-\right)$, follows the block form in Eq.~\eqref{eqSCX}, with $C$ a known invertible matrix. We choose symmetric invertible matrices $E_1$ and $E_2$ such that $C=E_1^{-1}E_2$, such a decomposition exists and is found efficiently \cite{taussky1972role}. We also define $F=E_1^{-1}+E_2^{-1}$. With this, $S\left(-\text{CX}-\right)$ is given by (see step-by-step realization in the SM \cite{SM}),
\begin{equation}
	S\left(-\text{CX}-\right)=
	\begin{bmatrix}
		I_n & 0 \\
		F C^T  & I_n
	\end{bmatrix}
	\begin{bmatrix}
		I_n & E_1 \\
		0 & I_n
	\end{bmatrix}
	\begin{bmatrix}
		I_n & 0 \\
		F & I_n
	\end{bmatrix}
	\begin{bmatrix}
		I_n & E_2 \\
		0 & I_n
	\end{bmatrix}.\label{eqMagic}
\end{equation}

Crucially, $F$ is a sum of two symmetric matrices and therefore symmetric. Moreover $F C^T=E_1^{-1}E_2 E_1^{-1}+E_1^{-1}$ is symmetric as well (derived in the SM \cite{SM}), thus
\begin{equation}
\begin{split}
	\text{CX}=& \umqz\left(F C^T\right)\umqx\left(E_1\right) \umqz\left(F\right)\\
	&  \umqx\left(E_2\right)e^{i\frac{\pi}{2}{\sum_k\eta_kX_k }} e^{i\frac{\pi}{2}{\sum_k\mu_kZ_k } },\label{eqMagic2}
\end{split}
\end{equation}
where the coefficients $\mu_k,\eta_k \in \{0,1\}$ are determined by the symplectic phase vector associated with the  $-\text{CX}-$ layer and the gates on the right hand-side of Eq.~\eqref{eqMagic}. Clearly an analogous decomposition with a leading $\umqz$ and final $\umqx$ gates exists and derived in a similar manner. 

Combining Eq. \eqref{eqMagic2} with the Clifford decomposition used in Eq.~\eqref{eqProctor} results in a five multiqubit gate realization. However, with the decomposition in Eq.~\eqref{eqDecomp} the leading and trailing layers are merged with the $\text{R}_{ZZ}$ and $\text{R}_{XX}$ layers, constituting a realization with up to four multiqubit gates.

For a realization in trapped ions qubits, the leading contribution to the drive power that is necessary to implement the gates in Eq.~\eqref{eqMagic2} can be estimated. This is an important quantity, as many gate-errors naturally scale with the drive power, e.g. photon scattering. It has been shown that the power required to drive $\umqz\left(\xi\right)$ scales as $\text{nuc}\left(\xi\right)$, the nuclear-norm of $\xi$, i.e. the $L_1$ norm of its eigenvalues \cite{shapira2023fast}. 

\begin{figure}
	\centering
	\includegraphics[width=0.8\textwidth]{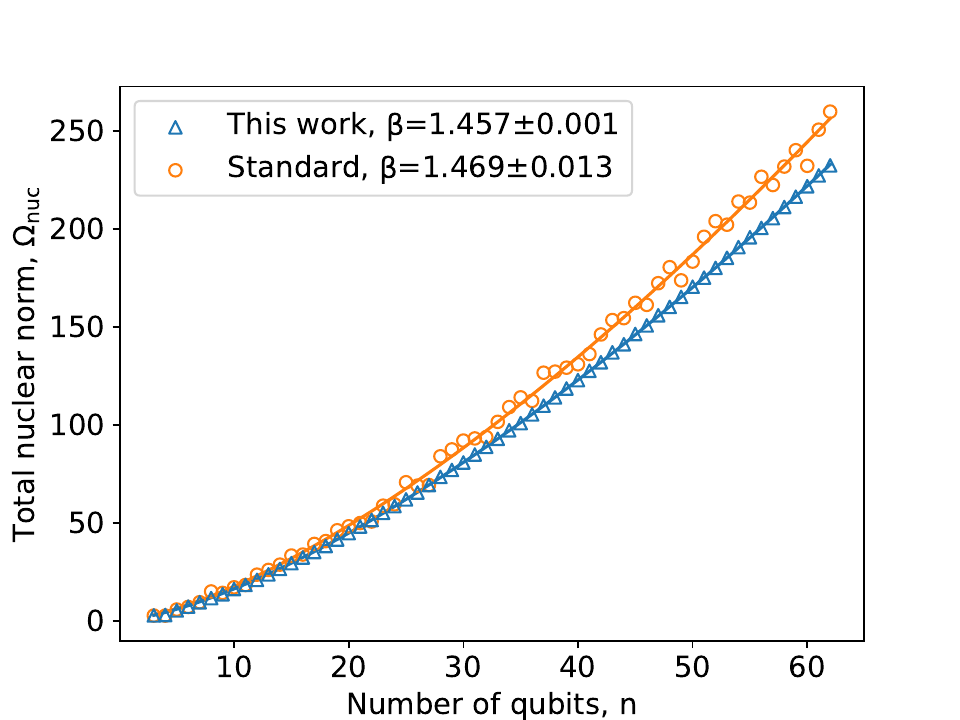} % no extension if PDF/PNG
	\caption{Total nuclear norm (vertical) of decompositions of random $-\text{CX}-$ layers over a varying number of qubits, $n$ (horizontal), with standard Gaussian elimination (orange) and with our constant-cost implementation (blue). Data (markers) are fitted to a power-law (solid) with $\Omega_\text{nuc}\propto n^\beta$, with the values of $\beta$ and confidence intervals shown in the legend.}
	\label{fig:nuclear}
\end{figure}

Thus, we consider the total drive power, $\Omega_\text{nuc}$, required to realize arbitrary classical reversible circuit, $-\text{CX}-$, over a varying number of qubits, $n$. This is compared to the total nuclear norm of a standard realization using Gaussian elimination. Figure \ref{fig:nuclear} shows the results of this analysis, with the total power of both methods fitted to a power law, $\Omega_\text{nuc}\propto n^\beta$. As seen, both method scale approximately as $n^{3/2}$ with our constant-cost implementation having a slightly reduced total nuclear norm.

In conclusion, we have shown a decomposition of Clifford operations at a constant-cost of, at most, four multiqubit gates, based on an implementation of a classical reversible circuit with, at most, four multiqubit gates. This gate count is optimal. We have further investigated the drive power associated with this implementation, showing that it can be achieved with essentially the same power as standard methods that utilize two-qubit gates, that is, circuit depth is significantly reduced without incurring drive power cost. 

\bibliography{references}

\cleardoublepage
\onecolumngrid
{
	\begin{center}
	{\large \bfseries Supplemental material \par}
	\end{center}	
	\bigskip
	\setcounter{section}{0}
	{
\section{Step-by-step derivation of Eq. (\ref{eqMagic})}

The symplectic matrix of a general CNOT circuit operating on $n$ qubits has the form,
$
	S_{\text{CX}}=
	\begin{bmatrix}
		C^{-T} & 0 \\
		0 & C
	\end{bmatrix}
$ \cite{nemirovsky2025reducedconstantcostimplementationsclifford}.
For every $C\in \mathbb{F}_2^{n\times n}$ one can always find symmetric matrices $S_1, S_2 \in \mathbb{F}_2^{n\times n}$ such that $ C=S_1 S_2 $ \cite{taussky1972role}. 
As, $C$ is invertible we define $E_1 = S^{-1}_1 $ and  $E_2 = S_2 $ such that $C=E_1^{-1}E_2$, as in the main text. Furthermore, we define $F=E_1^{-1}+E_2^{-1}$ and $G=F C^T$.

We note that $F$ being a sum of symmetric matrices is symmetric. We also note that, $G=F C^T=(E_1^{-1}+E_2^{-1})E_2 E_1^{-1} = E_1^{-1} E_2 E_1^{-1} + E_1^{-1}$. Hence $G$ is also symmetric.  

Then, since, 
\begin{equation}
E_2+E_1+E_1 F E_2=E_2+E_1+E_1 (E_1^{-1}+E_2^{-1} ) E_2=E_2+E_1+E_2+E_1=0
\end{equation}
It follows that,
\begin{align}
		\begin{bmatrix}
		I & 0 \\
		G & I
	\end{bmatrix}
		\begin{bmatrix}
		I & E_1 \\
		0 & I
	\end{bmatrix}
		\begin{bmatrix}
		I & 0 \\
		F & I
	\end{bmatrix}
	\begin{bmatrix}
		I & E_2 \\
		0 & I
	\end{bmatrix} & =
	\begin{bmatrix}
		I+E_1 F &  E_2+E_1+E_1 F E_2 \;\;\;\;\;\;  \\
		G+F +G E_1 F & \;\;\;\;\;\; G (E_2+E_1+E_1 F E_2 )+I+F E_2
	\end{bmatrix} = \nonumber \\
	& = 
		\begin{bmatrix}
		I+E_1 F & 0 \;\;\;\;\;\;  \\
		G+F +G E_1 F & \;\;\;\;\;\; I+F E_2
	\end{bmatrix},
\end{align}
and since,
\begin{equation}
	I+E_1 F=I+E_1 (E_1^{-1}+E_2^{-1} )=E_1 E_2^{-1}=C^{-T},
\end{equation}
we see that,
\begin{equation}
	\begin{bmatrix}
		I & 0 \\
		G & I
	\end{bmatrix}
	\begin{bmatrix}
		I & E_1 \\
		0 & I
	\end{bmatrix}
	\begin{bmatrix}
		I & 0 \\
		F & I
	\end{bmatrix}
	\begin{bmatrix}
		I & E_2 \\
		0 & I
	\end{bmatrix} =
	\begin{bmatrix}
		C^{-T} & 0 \;\;\;\;\;\;  \\
		G+F +G E_1 F & \;\;\;\;\;\; I+F E_2
	\end{bmatrix}.
\end{equation}
Next we obtain,
\begin{equation}
 I+F E_2=I+(E_1^{-1}+E_2^{-1} ) E_2=E_1^{-1} E_2=C,
\end{equation}
thus,
\begin{equation}
	\begin{bmatrix}
		I & 0 \\
		G & I
	\end{bmatrix}
	\begin{bmatrix}
		I & E_1 \\
		0 & I
	\end{bmatrix}
	\begin{bmatrix}
		I & 0 \\
		F & I
	\end{bmatrix}
	\begin{bmatrix}
		I & E_2 \\
		0 & I
	\end{bmatrix} =
	\begin{bmatrix}
		C^{-T} & \;\;\;\;\;\; 0   \\
		G+F +G E_1 F & \;\;\;\;\;\; C
	\end{bmatrix}.
\end{equation}

Finally, since, 
\begin{equation}
 G=F C^T=F E_2 E_1^{-1}=(E_1^{-1}+E_2^{-1} ) E_2 E_1^{-1},
\end{equation}
we obtain,
\begin{align}
G+F+G E_1 F= & F C^T+(E_1^{-1}+E_2^{-1} )+G E_1 F= 
\nonumber \\
= & (E_1^{-1}+E_2^{-1} ) E_2 E_1^{-1}+(E_1^{-1}+E_2^{-1} )+(F E_2 E_1^{-1})E_1 (E_1^{-1}+E_2^{-1} )= 
\nonumber \\
  = &(E_1^{-1} E_2 E_1^{-1}+E_1^{-1} )+(E_1^{-1}+E_2^{-1} )+F E_2 (E_1^{-1}+E_2^{-1} )= \nonumber\\
  = & E_1^{-1} E_2 E_1^{-1}+E_2^{-1}+F (E_2 E_1^{-1}+I)=E_1^{-1} E_2 E_1^{-1}+E_2^{-1}+(E_1^{-1}+E_2^{-1} )(E_2 E_1^{-1}+I) = \nonumber \\
  = & E_1^{-1} E_2 E_1^{-1}+E_2^{-1}+E_1^{-1} E_2 E_1^{-1}+E_1^{-1}+E_1^{-1}+E_2^{-1}=0,
\end{align}
and we conclude that indeed,
\begin{equation}
	\begin{bmatrix}
		I & 0 \\
		G & I
	\end{bmatrix}
	\begin{bmatrix}
		I & E_1 \\
		0 & I
	\end{bmatrix}
	\begin{bmatrix}
		I & 0 \\
		F & I
	\end{bmatrix}
	\begin{bmatrix}
		I & E_2 \\
		0 & I
	\end{bmatrix} =
	\begin{bmatrix}
		C^{-T} & 0   \\
		0 &  C
	\end{bmatrix} = S_{\text{CX}}.
\end{equation}

}
	}

\end{document}